\documentclass[twoside]{article}

\usepackage{fleqn,espcrc2}
\usepackage{graphicx}

\newcommand{\rmd}{{\rm d}}
\newcommand{\rmi}{{\rm i}}

\newcommand{\beq}{\begin{equation}}
\newcommand{\eeq}{\end{equation}}
\newcommand{\bea}{\begin{eqnarray}}
\newcommand{\eea}{\end{eqnarray}}

\newcommand{\lsr}{\langle n_{3} \rangle}
\newcommand{\llr}{\langle\lambda\rangle}

\newcommand{\lanthan}{La$_{2}$CuO$_{4}$ }
\newcommand{\ybco}{YBa$_{2}$Cu$_{3}$O$_{6.15}$ }

\title{Sublattice Magnetization and N\'eel Transition in the $2D$ Quantum 
Heisenberg Antiferromagnet}

\author{Eduardo C. Marino and Marcello B. Silva Neto 
\address{Instituto de F\'\i sica, Universidade Federal do Rio de Janeiro,
         Caixa Postal 68528, Rio de Janeiro - RJ, 21945-910, Brazil}
        \thanks{Research partially supported by CNPq and FAPERJ.}
        }

\begin{document}

\begin{abstract}

We present an analytic expression for the finite temperature sublattice magnetization,
at the Josephson scale, in two-dimensional quantum antiferromagnets with short range N\'eel 
order. Our expression is able to reproduce both the qualitative behaviour of the phase 
diagram $M(T)\times T$ and the experimental values of the N\'eel temperature $T_{N}$ for 
either doped YBa$_{2}$Cu$_{3}$O$_{6.15}$ and stoichiometric La$_{2}$CuO$_{4}$ compounds. 

\end{abstract}

\maketitle

\begin{figure}
\centerline{\includegraphics[width=2.9in]{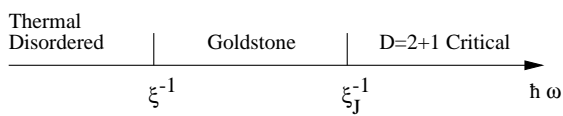}}
\vspace{-0.3in}
\caption{Properties of the $2D$ quantum Heisenberg antiferromagnet as
a function of the observation wave vector $k$, or frequency $\omega$.}
\vspace{-0.1in}
\label{Fig-Ren-Class}
\end{figure}

It is the purpose of this work to show that the experimental data for the 
sublattice magnetization of \lanthan \cite{Keimer} and \ybco \cite{Tranquada}
can in fact be described still in the context of a two-dimensional square-lattice 
quantum Heisenberg antiferromagnet at finite temperatures. Our starting point is the 
observation that the nature of the spin correlations in the renormalized classical 
regime is consistent with one of the three possibilities of fig. \ref{Fig-Ren-Class}, 
according to the observation wave vector \cite{Sachdev}. Next we argue that
the spin dynamics in the intermediate Goldstone region can be described by an effective
field theory for the low energy, long wavelength fluctuations of the spin fields about
a state with short range N\'eel order. In fact, since at low $T$ the three regions of fig. 
(\ref{Fig-Ren-Class}) are well separated, dynamic scaling hypotesis is valid and a 
hydrodynamic picture in which short wavelength spin waves follow adiabatically the 
disordered background is applicable \cite{Chakravarty}. Finally, we show that
the destruction of antiferromagnetic order in real materials can be associated with
the collapse of the Goldstone region in fig. (\ref{Fig-Ren-Class}). 

%
%

We start already in the continuum formulation of the problem by considering the partition 
function
\beq
{\cal Z}(\beta)=\int{\cal D}n_{l}{\;}\delta(n_{l}^{2}-1){\;}
\exp{(-{\cal I}(n_{l}))},
\label{Part-Func}
\eeq
where
$$
{\cal I}(n_{l})=\frac{\rho_{0}}{2\hbar}
\int_{0}^{\hbar\beta}\rmd\tau\int\rmd^{2}{\bf x}
\left[(\nabla n_{l})^{2}+
\frac{1}{c_{0}^{2}}(\partial_{\tau} n_{l})^{2}
\right] \nonumber
\label{Action}
$$
with $n_{l}=(\sigma,\vec{\pi})$, $l=1,\dots,N=3$. Let us next construct an effective field 
theory for the low frequency, long wavelength fluctuations of the staggered components of 
spin-fields about a state with short range N\'eel order. To this end we perform integration 
of the Fourier components of the fields in (\ref{Part-Func}) with frequency inside momentum 
shells $\kappa\leq |\vec{k}| \leq\Lambda$. The resulting partition function is such that, for 
large $N$, the leading contribution comes from the scale dependent stationary configurations 
$\lsr_{\kappa}$ and $\rmi\llr_{\kappa}=m_{\kappa}^{2}$, solutions of the saddle-point equation
\beq
\lsr_{\kappa}^{2}=
\frac{1}{g_{0}}-\frac{1}{\beta}
\sum_{n=-\infty}^{\infty}\int_{\kappa}^{\Lambda}
\frac{\rmd^{2}{\bf k}}{(2\pi)^{2}}
\frac{1}{{\bf k}^{2}+\omega_{n}^{2}+m_{\kappa}^{2}},
\label{New-Saddle-Point-Eq}
\eeq
where, as usual, $\lambda$ is a Lagrange multiplier for the averaged fixed length constraint 
and $g_{0}=N/\rho_{0}$.

Due to the presence of the IR cutoff $\kappa$, the system can be found in two different phases 
(regimes): ordered (asymptotically free) or disordered (strongly coupled); depending on the 
size of $\xi_{\kappa}=1/\kappa$ relative to $\xi$: smaller (high energies) or larger 
(low energies). In the ordered phase, $\xi_{\kappa}\ll\xi$, $m_{\kappa}=0$ is the solution 
that minimizes the free energy and the $2D$ system is then characterized by a nonvanishing 
effective sublattice magnetization $\lsr_{\kappa}\neq 0$, a divergent effective correlation 
length $\xi_{eff}=1/m_{\kappa}=\infty$ and gapless excitations in the spectrum. 

We can subtract the linear divergence in (\ref{New-Saddle-Point-Eq}) by the renormalization
$1/g_{0}=1/g_{c}+\rho_{s}/4\pi N$, where $g_{c}=4\pi/\Lambda$ and 
$\rho_{s}=(\hbar c/a)\;\sqrt{S(S+1/2)}/2\sqrt{2}$ is the bulk spin stiffness. Now, after 
momentum integration and frequency sum, we obtain the running spin stiffness
\beq
\rho_{s}(\kappa,\beta)=\frac{\rho_{s}}{2}+\frac{N}{\beta}\ln{(2\sinh{(\beta\kappa/2)})}.
\label{Running-rho}
\eeq
It is clear from the above expression that at long distances, $k\rightarrow 0$, the system 
is found disordered and strongly coupled. No sublattice magnetization can be measured. Here, 
conversely, in order to obtain a finite temperature phase transition in the $2D$ system, we 
will rather fix the scale $\kappa$ and study the behaviour of the spin stiffness 
(\ref{Running-rho}) with the running parameter as being the temperature. For this it suffices 
to impose the boundary condition 
\beq
\rho_{s}(\kappa,T=0)=\rho_{s}.
\label{Bound-Cond}
\eeq  
From (\ref{Bound-Cond}) we conclude that $\kappa=\rho_{s}/N$, which is exactly the
inverse Josephson correlation length $\kappa=\xi_{J}^{-1}$. This should not be surprising 
since the spin stiffness is itself a microscopic, short wavelength quantity defined at the 
Josephson scale. Now, inserting (\ref{Bound-Cond}) in (\ref{Running-rho}), the expression 
for the finite temperature effective sublattice magnetization, 
$M(T)\equiv\rho_{s}(\kappa=\rho_{s}/N,T)$, becomes 
\beq
M(T)=\frac{M_{0}}{2}+NT
\ln{\left(2\sinh{\left(\frac{M_{0}}{2NT}\right)}\right)},
\label{Magnetization}
\eeq
with $M_{0}=\rho_{s}$. As the temperature increases the sublattice magnetization 
$M(T)$ vanishes at a N\'eel temperature $T_{N}$ given by
\beq
T_{N}=\frac{M_{0}}{N\ln{2}}=\frac{\xi_{J}^{-1}}{\ln{2}}.
\label{Critical-Temperature}
\eeq
This also corresponds to the value of $T$ for which $\xi=\xi_{J}$ and
the Goldstone region in fig. (\ref{Fig-Ren-Class}) collapses.

\begin{figure}
\centerline{\includegraphics[width=2.0in]{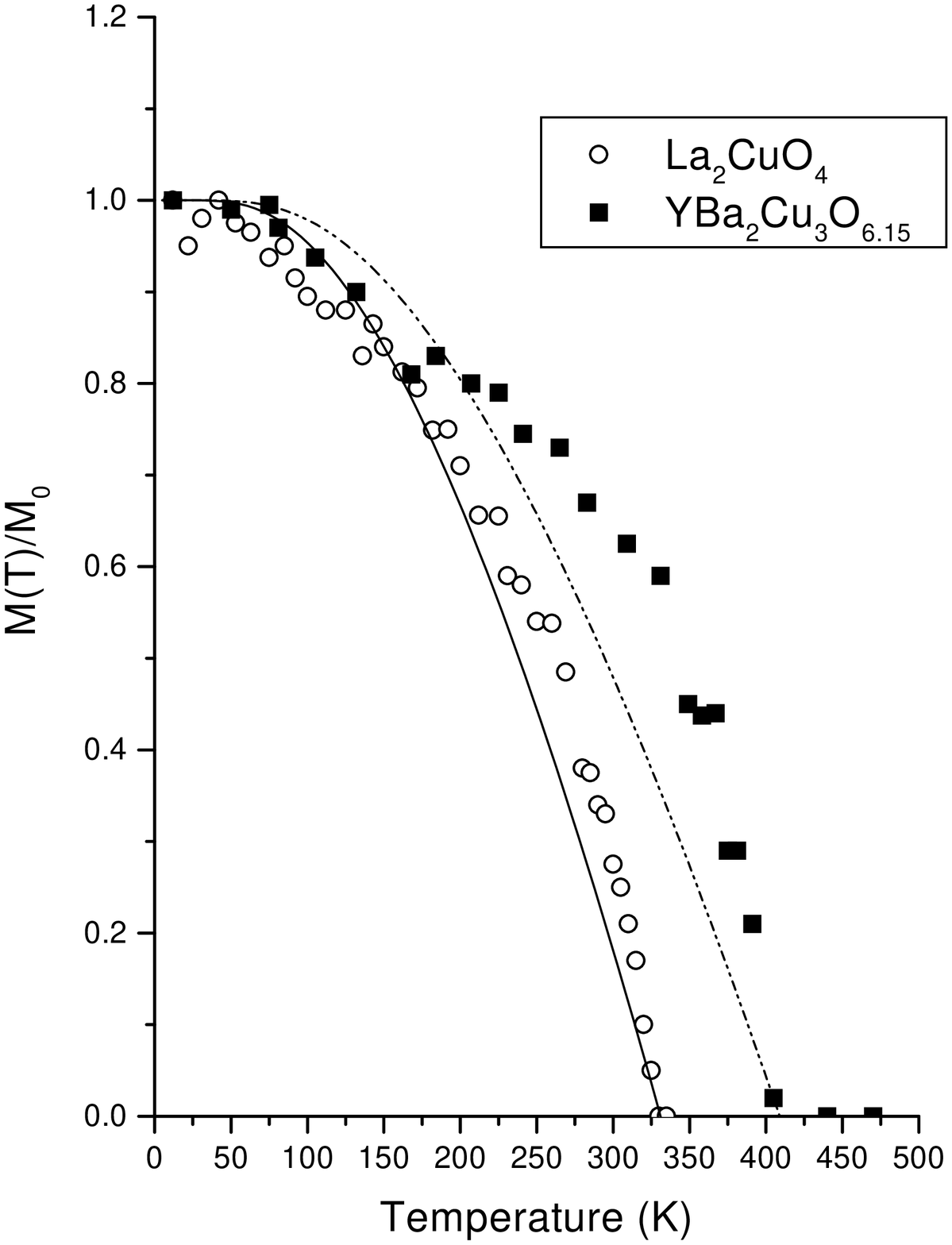}}
\vspace{-0.3in}
\caption{Relative sublattice magnetization at the Josephson scale for
\lanthan (solid line and $\hbar c = 0.78$ eV $\AA$) and \ybco (dash dot dot line and 
$\hbar c=0.90$ eV $\AA$).}
\vspace{-0.1in}
\label{Fig-Compounds}
\end{figure}

In fig. (\ref{Fig-Compounds}) we plot $M(T)/M_{0}$ against experiment for either \lanthan 
and \ybco and for $N=3$. To compute $\rho_{s}$ we have used $S=1/2$, $a=3.8 \AA$ and the
experimental values of $c$. A more detailed analysis of the problem, with a discussion on 
how the actual deviation of the experimental points from the theoretical curves can be 
accounted for, can be found in \cite{Marino-Marcello}.

%
%

\end{document}